\documentclass[preprint,12pt]{elsarticle}

\usepackage{amssymb}

\usepackage{graphicx}
\usepackage{caption}
\usepackage{makecell}
\usepackage{amsmath}
\usepackage{algorithm}
\usepackage{listings}
\usepackage{xcolor}
\usepackage{color}
\usepackage[noend]{algpseudocode}
\usepackage{multirow}
\usepackage{subcaption,siunitx,booktabs}
\usepackage[margin=1in]{geometry}
\usepackage{comment}

\journal{The Journal of Systems and Software}

\begin{document}

\begin{frontmatter}


\title{Identifying Algorithm Names in Code Comments}



\author[label1]{Jakapong Klainongsuang}
\author[label2]{Yusuf Sulistyo Nugroho}
\author[label2]{Hideaki Hata}
\author[label1]{Bundit Manaskasemsak}
\author[label1]{Arnon Rungsawang}
\author[label1]{Pattara Leelaprute}
\author[label2]{Kenichi Matsumoto}
\address[label2]{Nara Institute of Science and Technology, Japan}
\address[label1]{Kasetsart University, Thailand}


\begin{abstract}
For recent machine-learning-based tasks like API sequence generation, comment generation, and document generation, large amount of data is needed.
When software developers implement algorithms in code, we find that they often mention algorithm names in code comments.
Code annotated with such algorithm names can be valuable data sources.
In this paper, we propose an automatic method of algorithm name identification. The key idea is extracting important N-gram words containing the word `algorithm' in the last. We also consider part of speech patterns to derive rules for appropriate algorithm name identification. The result of our rule evaluation produced high precision and recall values (more than 0.70). We apply our rules to extract algorithm names in a large amount of comments from active FLOSS projects written in seven programming languages, C, C++, Java, JavaScript, Python, PHP, and Ruby, and report commonly mentioned algorithm names in code comments.
\end{abstract}


\end{frontmatter}


\section{Introduction}
\label{sec:introduction}

The importance of annotated data is increasing for various machine-learning-based tasks, such as generating API sequences~\cite{Gu:2016:DAL:2950290.2950334}, comment generation~\cite{Jiang:2017:AGC:3155562.3155583}, pseudo-code generation~\cite{Oda:2015:LGP:2916135.2916173}, and document generation~\cite{Wong:2013:AMQ:3107656.3107727,8530113}.
For preparing annotated data, manual annotation is considered to be an effective approach for high quality data (for example, collecting code and comment pairs~\cite{Oda:2015:LGP:2916135.2916173}). However, because of required human effort, increasing annotated data is costly and not easy. For automating preparation of annotated data, Yin et al. proposed a method to mine aligned code and natural language pairs from Stack Overflow~\cite{Yin:2018:LMA:3196398.3196408}.

In this paper, we focus on algorithms as annotations of code.
In the field of mathematics and computer science, algorithms denote one of the main subjects of study that are used to processing a set of input data to obtain the output, operating computation, or even automatically finishing the task of reasoning\footnote{https://en.wikipedia.org/wiki/Algorithm}. 
Indeed, they are very essential due to their problem-independence but obscure in control strategy applications, or the programming language allocation as well \cite{SMITH1990305}.
Toward automatically preparing code annotated with algorithm names, this paper addresses identifying algorithm names in code comments. To the best of our knowledge, this is the first study of mining algorithm names from code comments.

\begin{small}
    \begin{lstlisting}[caption=A code comment containing an algorithm name (Insertion Sort algorithm), label=code:examplecodecomment]
    Sorts the array of size count using comparator lessThan using an Insertion 
    Sort algorithm.
    \end{lstlisting}
\end{small}

To explain the design of implemented code, we find that developers sometimes explicitly mention algorithm names in the associated code.
Listing \ref{code:examplecodecomment} shows an example of the algorithm provision inline with a code comment\footnote{https://github.com/servo/skia/blob/v0.30000019.0/src/core/SkTSort.h\#L95}. The comment mentions the name of the algorithm, \textit{insertion Sort algorithm}.
%
%



In this paper, we propose a method of identifying algorithm names and report the frequencies of algorithm names appear in code comments of free/libre and open source software (FLOSS).
The key idea of our identification is extracting important N-gram words containing the word `algorithm' in the last. 
We also consider part of speech patterns to derive rules for appropriate algorithm name identification. 
The result of our rule evaluation produced high precision and recall values (more than 0.70). 
We apply our rules to extract algorithm names in a large amount of comments from active FLOSS projects written in seven programming languages, C, C++, Java, JavaScript, Python, PHP, and Ruby, and report commonly mentioned algorithm names in code comments.

Our contributions can be summarized as follows.
\begin{itemize}
    \item We propose a method to identify algorithm names in code comments.
    \item Proposed method is manually evaluated with preliminary data containing 1,581 identified N-gram terms and 458 summarized distinct N-gram terms.
    \item By applying our method to large-scale FLOSS data, we report commonly used algorithms in for seven languages (C, C++, Java, JavaScript, Python, PHP, and Ruby).
\end{itemize}

\section{Algorithm Name Identification}

\subsection{Overview}
\begin{figure}[ht]
    \centering
    \includegraphics[width=\textwidth]{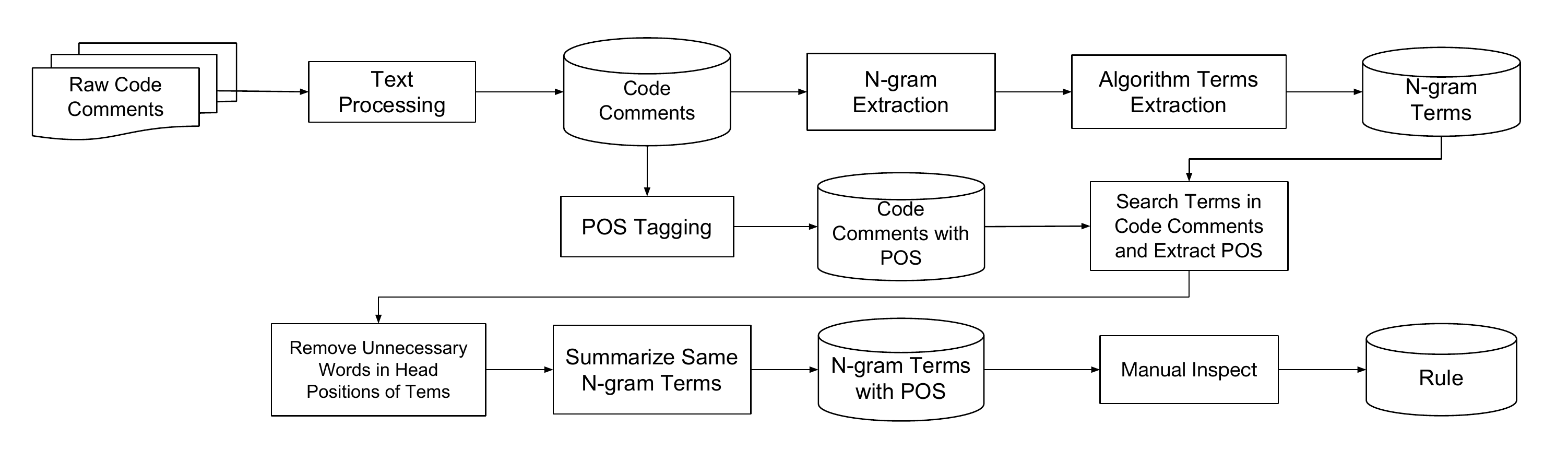}
    \caption{Overview of our FLOSS in creating the rule}
    \label{fig:approachoverview}
\end{figure}

In this section, we give an overview of our proposed method.
Fig. \ref{fig:approachoverview} shows the main steps of our method. In this study, we obtained the preliminary code comment data from FLOSS projects hosted in GitHub to create rules to identify the algorithm names.

From code comments, we extract word terms using N-gram IDF similar to the study~\cite{Terdchanakul:8094457}. First, to identify the algorithm name, we obtain N-gram terms containing keywords `algorithm' in the last position (e.g. \textit{quick sort algorithm, search algorithm}). Second, the part of speech (POS) tagging process is applied for all code comments. 
Finally, we subsequently remove the unnecessary words in head position of the N-gram word terms by considering POS tags.
With the above process, We create \textit{Inclusive} and \textit{Exclusive} criteria to identify appropriate algorithm names.

As the processes show, we do not target algorithm names that do not include `algorithm' in the bottom. We consider this makes our identification method precise by ignoring inappropriate terms including words similar to algorithm names. In addition, our method do not need a list of algorithm names, which make our method robust.

\subsection{Text Preprocessing, N-gram Extraction, and POS tagging}
To create our inclusive and exclusive criteria, we use our preliminary data of
code comments containing the keywords of the algorithms from two repositories of C programming language (i.e. \texttt{gecko-dev} and \texttt{linux}). Special characters such as `*', `\#',  `/' were removed and the part of speech to all code comments were tagged using the spacy library~\cite{Omran:2017:CNL:3104188.3104213}.

We applied N-gram IDF, a
theoretical extension of Inverse Document Frequency (IDF) introduced by Shirakawa~\cite{
Shirakawa:2015:NIG:2736277.2741628,Shirakawa:2017:IWN:3077622.3052775}, to capture N-gram terms and obtained terms that contain `algorithm' in the last position. Each N-gram word terms were then searched in all code comments. If they match the words in each comment, they will be tagged with the same part of speech.

\subsection{Remove Unnecessary Words and Summarize Same N-gram Terms}
In this step, we recursively remove the unnecessary words shown in Table \ref{tab:unnecessarywords} in the head position, as those words cannot be algorithm names. 
If in a single comment contains the terms of ``quick sort algorithm'', we can extract both sort algorithm and quick sort algorithm. In this case we select the longest one, ``quick sort algorithm''. 

\begin{table}[ht]
    \caption{Unnecessary words in the head position}
    \label{tab:unnecessarywords}
    \centering
    \footnotesize
    \begin{tabular}{l|l}
        \hline
        \textbf{Part of speech} & \textbf{Description}  \\
        \hline
        VERB ADP & verb with conjunction, subordinating or preposition  \\
        ADP & conjunction, subordinating or preposition \\
        NUM & number    \\
        DET & determine \\
        \hline
    \end{tabular}
\end{table}

Because the same N-gram word terms can be tagged in different part of speech depend on the position of the words in each code comment, we summarize them by the majority of part of speech. Table \ref{tab:findmajorityofngram} shows ``sort algorithm'' from four different code comments. There are three terms tagged as NOUN NOUN while the last one is tagged as VERB NOUN. Because the sorting algorithm has a majority of 3 out of 4 candidate parts of speech, thus we set the majority of sort algorithm as NOUN NOUN. Meanwhile, the term ``blur algorithm'' has no majority in this case, 
we ignore such algorithm name candidates.

\begin{table}[ht]
    \caption{Example of N-gram terms and their part of speech}
    \label{tab:findmajorityofngram}
    \centering
    \footnotesize
    \begin{tabular}{c|c}
        \hline
        \textbf{N-gram term} & \textbf{Part of speech}\\
        \hline
        sort algorithm & NOUN NOUN\\
        sort algorithm & NOUN NOUN\\
        sort algorithm & NOUN NOUN\\
        sort algorithm & VERB NOUN\\
        \hline
        blur algorithm & NOUN NOUN\\
        blur algorithm & ADJ NOUN\\
        \hline
    \end{tabular}
\end{table}

\subsection{Creating Rules}
\label{sec:rulecreating}
In this step, we classify the N-gram word terms using POS tag and manually create identification rules. We focus only on the reliable N-gram word terms; terms that do not have majority nor the number of candidate is 1 are excluded. We manually verified correct algorithm names or not for all obtained algorithm name candidates in the preliminary data and labeled \textit{valid} or \textit{invalid}.

If the number of valid algorithm name in the part of speech is larger than the invalid one, it will be included in the Inclusive rule sets. Otherwise, if the number of valid algorithm name in the part of speech is lesser than the invalid, this will be considered to be an Exclusive rule. Inclusive rule means the unlabeled term is marked as a valid algorithm name, 
while the exclusive rule means the unlabeled term is assigned as an invalid algorithm name. 


Next, we build a program to implement the rule automatically with bigger data. The procedure we used in the code is shown in Algorithm \ref{alg:algorithmnamedetectingrule} with the all created rules. Description of part of speech is shown in Table \ref{tab:description_of_POS}.

\begin{table}[ht]
    \caption{Description of Part of Speech}
    \label{tab:description_of_POS}
    \centering
    \footnotesize
    \begin{tabular}{l|p{5.2cm}}
        \hline
        \textbf{Part of speech} & \textbf{Description}  \\
        \hline
        VERB & verb \\
        VERB(-ing) & verb end with suffix ``ing''\\
        NOUN & noun    \\
        PART & particle \\
        CONJ & conjunction, coordinating \\
        ADJ & adjective \\
        ADV & adverb\\
        \hline
    \end{tabular}
\end{table}



\begin{algorithm}
\caption{Algorithm Name Identification}
\label{alg:algorithmnamedetectingrule}
\footnotesize
\begin{algorithmic}[1]
\Procedure{rule checking (part of speech of unlabeled term)}{}
\State $\textit{POS} \gets \text{part of speech of }\textit{unlabeled term}$
\If {$POS \text{ is included }  \textit{only NOUN}$} 
\State \Return valid
\ElsIf{$POS \text{ is included }  \textit{CONJ}$}
\State \Return invalid
\ElsIf{$POS \text{ is included }  \textit{DET}$}
\State \Return invalid
\ElsIf {$POS \text{ is included }  \textit{VERB}$} 
    \If {$POS = \textit{VERB(-ing) NOUN}$} 
    \ElsIf {$POS = \text{ NOUN VERB NOUN }$} 
    \State \Return valid
    \ElsIf {$POS = \text{ VERB ADJ NOUN NOUN NOUN }$} 
    \State \Return valid
    \EndIf
    \State \Return invalid
\ElsIf{$POS \text{ is included }  \textit{only ADJ and NOUN}$}
    \If {$POS = \textit{ADJ ADJ NOUN}$} 
    \State \Return valid
    \ElsIf {$POS = \textit{ADJ NOUN NOUN}$} 
    \State \Return valid
    \ElsIf {$POS = \textit{NOUN ADJ NOUN}$} 
    \State \Return valid
    \ElsIf {$POS = \textit{ADJ ADJ NOUN NOUN}$} 
    \State \Return valid
    \ElsIf {$POS = \textit{ADJ NOUN ADJ NOUN}$} 
    \State \Return valid
    \EndIf
    \State \Return invalid
\ElsIf{$POS \text{ is included }  \textit{ADP}$}
    \If {$POS = \textit{ADV ADP NOUN}$} 
    \State \Return valid    
    \EndIf
    \State \Return invalid
\ElsIf{$POS \text{ is included }  \textit{ADV}$}
    \If {$POS = \textit{ADV NOUN}$} 
    \State \Return valid  
    \ElsIf{$POS = \textit{ADV PART NOUN}$} 
    \State \Return valid 
    \ElsIf{$POS = \textit{ADV ADJ ADJ NOUN}$} 
    \State \Return valid 
    \ElsIf{$POS = \textit{ADV ADJ ADJ NOUN NOUN}$} 
    \State \Return valid 
    \EndIf
    \State \Return invalid
\EndIf
\EndProcedure
\end{algorithmic}
\end{algorithm}

\section{Identification Evaluation with Preliminary Data}
\label{sec:evaluation}
In this section, we evaluate our method shown in Algorithm \ref{alg:algorithmnamedetectingrule} with the preliminary data. 
One of the authors manually created an oracle of all algorithm name candidates with valid and invalid labels. All algorithm name candidates in the preliminary data are classified into valid or invalid with our method and are compared with the oracle.
We present Precision, Recall, and F-measure in Table \ref{tab:evaluationscore}.
Precision measures accuracy of Algorithm \ref{alg:algorithmnamedetectingrule} to to correctly identify valid algorithm names. Recall is the fraction of the valid algorithm name that are successfully retrieved. F-measure is the harmonic mean of Precision and Recall. We obtained considerably high accuracy (more than 0.7 in Precision, Recall, and F-measure).

\begin{table}[ht]
    \caption{Performance of algorithm name identification}
    \label{tab:evaluationscore}
    \centering
    \footnotesize
    \begin{tabular}{ll}
        \hline
        \textbf{Metric} & \textbf{Score}\\
        \hline
        Precision & 0.76\\
        Recall & 0.70\\
        F-measure & 0.73\\
        \hline
    \end{tabular}
\end{table}

\section{Applying to Large-scale FLOSS Data}
\label{sec:result}

There are some Web articles indicating that developers and students are interested in knowing important algorithms~\cite{web1,web2,web3}. Although there are such interests in understanding algorithm usages in practice, there is no empirical study for algorithm uses in FLOSS as far as we know.

To investigate the frequencies of algorithm names in FLOSS code comments,
we collect repositories in GitHub that have more than 500 commits in entire histories, and at least 100 commits in the most active two years, which is set to ignore long-term inactive projects and very short term active projects (university classes, for example). Code comments are extracted from those identified repositories: C 2,771, C++ 3,563, Java 4,995, JavaScript 7,130, Python 5,263, Ruby 2,233, and PHP 3,279.

Table \ref{tab:resultofmaindata} is the obtained top 10 results after applying our rules by frequency (the numbers below algorithm names). 
In the 70 algorithm names, identified 15 names were not appropriate algorithm names, such as ``learning algorithm'', ``following algorithm'', and ``legacy algorithm'', which is 0.786 (55/70) precision. Since we can immediately observe such inappropriate algorithm names, we exclude them in the results.


\begin{table}[ht]
    \caption{Top ranked algorithms found in seven programming languages. The term `algorithm' is abbreviated.}
    \label{tab:resultofmaindata}
    \centering
        \resizebox{\columnwidth}{!}{%
        \begin{tabular}{c|c|c|c|c|c|c|c}
            \hline\noalign{\smallskip}
            \multirow{2}{*}{\textbf{Rank}} & \multicolumn{7}{c}{\textbf{Programming Language}}	\\
            \cline{2-8}
            \noalign{\smallskip}
            & \textbf{Java} & \textbf{Ruby} & \textbf{Python} & \textbf{PHP} & \textbf{C} & \textbf{Cpp} & \textbf{JavaScript}	\\
            \noalign{\smallskip}\hline\noalign{\smallskip}
            1 & Search & Encryption & Nagles & Encryption & Hash & Compression & Ordering \\
            & 3,142 & 537 & 123 & 213 & 3,193 & 925 & 799 \\
            \hline
            2 & Optimization & Compression & Signature & RC4 & Compression & Unicode Bidirectional & Sort    \\
            & 493 & 22 & 119 & 206 & 2,592 & 711 & 599   \\
            \hline
            3 & Neighbour search & Audionormalizationsettings & SBO & Fragment parsing & Scheduling & Slow iterative & Diff    \\
            & 454 & 22 & 86 & 157 & 2,472 & 597 & 534    \\
            \hline
            4 & Estimation & Hash & Parsing & Search & Sfrcacc & Sorting & Parsing \\
            & 451 & 15 & 74 & 137 & 2,363 & 523 & 513    \\
            \hline
            5 & Signature & Search & Hashing & Depthfirst search & Search & Dct & Ritters   \\
            & 379 & 12 & 70 & 87 & 1,939 & 448 & 448 \\
            \hline
            6 & Sorting & Kex & Exchange & Ordering & Encryption & Hash & O(nd) Difference  \\
            & 377 & 12 & 67 & 84 & 1,774 & 428 & 448 \\
            \hline
            7 & Encryption & Serverside encryption & Encryption & Signing & Auth & MD5 & Search \\
            & 370 & 10 & 66 & 51 & 1,397 & 415 & 406 \\
            \hline
            8 & Binary search & Tarjans & Generation & Header canonicalization & Rate control & LLM & Clipping \\
            & 266 & 9 & 61 & 51 & 1,340 & 410 & 351  \\
            \hline
            9 & Stemming & Matching & Nagle & Body canonicalization & Public key & Euclids & Iteration   \\
            & 238 & 9 & 56 & 51 & 1,073 & 391 & 322  \\
            \hline
            10 & Hash & Signature & Matching & Canonicalization & Authentication & Clipping & Ray casting \\
            & 223 & 5 & 55 & 50 & 1,070 & 389 & 284  \\
            \hline\noalign{\smallskip}
        \end{tabular}%
        }
\end{table}

We see that words of
search algorithm appear frequently in Java, Ruby, PHP, C, and JavaScript. Encryption algorithms appear in Java, Ruby, Python, PHP, and C. 
In C programming language, most of top 10 algorithm related to security. Some group of algorithms such as search, parsing, hash, sorting are implemented in many programming languages. PHP and JavaScript programming language have some algorithm related to Web.

Table \ref{tab:example_results} shows example of code comments including identified algorithm names as well as programming languages, organization, repository, and file names. 
By extracting associated code with the identified algorithm names, it seems possible to automatically collect annotated code, in the future.

\begin{table*}[ht]
    \centering
    \caption{Example of Results} 
    \label{tab:example_results}
    \resizebox{\columnwidth}{!}{%
    \begin{tabular}{l|l}
    \hline
        \textbf{Language, Organization/Repository, Filename} & \textbf{Example of Comment} \\
    \hline
        C, & \textbf{/*Encryption Algorithm} for Unicast Packet */ \\
        dorimanx/Dorimanx-LG-G2-D802-Kerne, &    \\
        wifi.h &  \\
    \hline
        Ruby, & \# Instantiates one of the Transport::Kex classes (based on the negotiated  \\
        openshift/openshift-extras, & \# \textbf{kex algorithm}), and uses it to exchange keys. */   \\
        algorithms.rb & \\
    \hline
        Java, & // The reason for this method name, as opposed to getFirstStrongDir(), is that    \\
        codenameone/CodenameOne, & // ``first strong'' is a commonly used description of Unicode's \textbf{estimation algorithm}  \\
        BidiFormatter.java &    \\
    \hline
        C++, & /* This is the central step in the \textbf{MD5 algorithm} \\ 
        hwine/test-mc-ma-cvs, & \\
        md5.cc &    \\
    \hline
        Python, & \# Enable \textbf{Nagle's algorithm} for proxies, to avoid packet   fragmentation. \\
        Andersh75/resurseffektivitet, & \# We cannot know if the user has added default socket options, so we cannot replace the  \\
        connectionpool.py & \# list.    \\
    \hline
        PHP, & * Returns the input text encrypted using \textbf{RC4 algorithm} and the specified key.   \\
        DerDu/SPHERE-Framework, &  * RC4 is the standard encryption algorithm used in PDF format  \\
        tcpdf\_static.php &  \\
    \hline
        JavaScript, & // \textbf{ray casting algorithm} for detecting if point is in polygon \\
        CartoDB/d3.cartodb, & \\
        leaflet.js &    \\
    \hline
    \end{tabular}%
    }
\end{table*}

\section{Related Work}
\label{sec:relatedwork}


\subsection{Benefits of code comments}
Some studies were undertaken to analyze the importance of code comments in maintaining a software. According to Tenny \cite{Tenny:6171}, the comprehension of a programmer to acquire such important information of a source code relies on the readability of the program. This is very crucial in software maintenance. The more readable of a code, the easier the developers to maintain the program. The experiment was designed by comparing the effect of procedures and code comments tested to more than 100 software engineering students. From the analysis, it shows that code comments written by authors improve the ability to read the program. It is more significant if there are no procedures provide in the source code. Conversely, the procedures affect slightly to the readability of a program.

Similar work also performed by \cite{Woodfield:1981:EMC:800078.802534} in searching the advantage of comments in a code. In this study, the authors analyzed the relationship between the types of modularization and comments and the ability of programmers to comprehend the codes. Several questions were asked to some programmers based on four different modularization types (monolithic, functional, super, and abstract data type) of the same program with and without comments. The results of the observation indicate that programmers were easier to provide the answers if the source code contains comments. Furthermore, the version of modularization which makes the subjects able to perform better is the abstract data type.


\subsection{Measurement of code comments}
Hu et al. \cite{Hu:2018:DCC:3196321.3196334} investigated the effectiveness of the code comments used by developers. They argued that comments in the code are very essential to guide the programmers to understand the source code and make them easier and faster to analyze the programs. However, these comments are frequently incompatible, old-fashioned or even misplaced in a project. By applying the NLP on FLOSS projects, the authors propose an automatic tool to generate comments for methods written in Java. The finding shows that the proposed method surpasses the existing techniques with prominent divergent.

An analysis and assessment of code comment quality were also conducted by \cite{Steidl:6613836}. The authors state that software developers rely on their understanding of the source code in terms of development and maintenance. Nevertheless, their capabilities on the comprehension of the programs depend on how high the quality of the code comments. In the study, the FLOSS taken to address the problems consists of comments classification, quality model development, model assessment, and validity evaluation. The study indicates that the proposed method offers the analysis more detail compared to the existing techniques on the classification of code comment.

\section{Threats to Validity}
\label{sec:threatstovalidity}

Threats to the \textit{construct validity} exist in our rule creation approach with the preliminary data. Since rules were created only with the limited data, it may have limitations to generalize.
As seen in good precision from the result of large-scale data (Section~\ref{sec:result}), we consider that the rules are robust and not limited to specific projects.

Threats to the \textit{external validity} exist in our data preparation. Although we analyzed a large amount of repositories on GitHub, we cannot generalize our findings to industry nor open source projects in general; some of open source repositories are hosted outside of GitHub, e.g., on GitLab or private servers. Further studies are required.

\section{Conclusion}
\label{sec:conclusion}
In this paper, we have presented a method to identify algorithm names from code comment by creating rules using N-gram IDF and part of speech tagging. 
We find major part of speech of N-gram word terms by a majority of them. 
Our evaluation shows our method accurately identify algorithm names in FLOSS code comments. 
In the future, we plan to identify associated algorithm implementation code as well as algorithm names. 
Using such data, we could try helping developers by recommending actual code implementing algorithms from active FLOSS projects.

\section*{References}












\end{document}